\documentclass[a4paper,twoside,reqno]{bjp}
\usepackage{graphicx}
\usepackage{cite}
\usepackage{amssymb,amsmath,amscd,amsthm}
\usepackage{times}
\usepackage[bookmarks=false]{hyperref}
\hypersetup{%
	colorlinks=true,        
	linkcolor=blue,          
	citecolor=blue,         
	urlcolor=blue           
}
\usepackage{orcidlink}

\usepackage{geometry}
\allowdisplaybreaks

\begin{document}
	
	\title{Non-interacting String and Holographic Dark Energy Cosmological Models in $f(R)$ Theory of Gravitation}
	
	\runningheads{S. P. Hatkar, D. P. Tadas, A. S. Agrawal, S. D. Katore, }{Non-interacting String and Holographic Dark Energy Cosmological Models in $f(R)$ Theory}
	
	\begin{start}{%
			\author{S. P. Hatkar\orcidlink{0000-0002-3970-481X}}{1},
			\author{D. P. Tadas$^*$\orcidlink{0000-0002-1572-6213}}{2}
			\author{A. S. Agrawal\orcidlink{0000-0003-4976-8769}}{3},
			\author{S. D. Katore\orcidlink{0000-0003-0521-4334}}{4}
		
		\address{Department of Mathematics, A.E.S. Arts, Commerce and Science College, Hingoli-431513, India}{1}
		\address{Department of Mathematics, Toshniwal Arts, Commerce and Science College Sengaon, Dist. Hingoli, India}{2}
		\address{Department of Mathematics, Jayawant Shikshan Prasarak Mandal, University, Pune, India.}{3}
		\address{Department of Mathematics, Sant Gadge Baba Amravati University, Amravati-444602, India}{4}
		*Corresponding author’s E-mail: dtadas144@rediffmail.com
		
		\received{24 Feb 2024}
	}

	\begin{Abstract}
		In this paper, a new class of string and holographic dark energy (HDE) cosmological model in the context of the $f(R)$ theory of gravity using the Kasner metric is considered. The exact solution of the field equations is obtained using the relation between the average scale factor and the scalar function $f(R)$. It is observed that the universe is accelerating and expanding. The string phase of the universe is present at an early stage of the evolution of the universe. The universe is dominated by quintessence type HDE at present. The effect of the curvature function $f(R)$ is also observed on dynamical parameters.
	\end{Abstract}
	
	\begin{KEY}
		String, Holographic dark energy, Kasner metric, $f(R)$ gravity.
	\end{KEY}
\end{start}


\section{Introduction} 

Nowadays, it is widely accepted that the universe is experiencing an accelerated expansion. Recent observations from type Ia supernovae \cite{Riess1998, Perlmutter1999} combined with large-scale structure \cite{Tegmark2004a} and cosmic microwave background (CMB) \cite{Bennett2003, Spergel2003} anisotropies have provided the primary evidence for this cosmic acceleration. Two different approaches have been proposed to address the issue of accelerated expansion of the universe: the first one is the concept of dark energy, and the second is modifications to general relativity. As per the first approach, the accelerated expansion is caused by an exotic type of force known as "dark energy", whose exact nature is still unknown. Several dynamical dark energy (DE) candidates have been proposed to understand the nature of DE, including the cosmological constant, quintessence \cite{Ratra1998}, phantom \cite{Caldwell2002, Nojiri2003}, k-essence \cite{Chiba2000}, tachyon \cite{Sen2002} and interacting Chaplygin gas models \cite{Gorini2003}. The equation of state of dark energy (DE) is used to describe cosmic evolution and the accelerated expansion of the universe, and it is defined as $ \omega=\frac{p}{\rho} $, where $\rho$ and $ p $ are the energy density and pressure respectively.

The second approach is modification of GR, these theories are known as modified theories of gravitation. Among the several modified theories of gravitation, the $f(R)$ theory of gravity is regarded as the simplest and most straightforward generalization of Einstein's general theory of relativity (GTR). The higher-order curvature invariants are the functions of the Ricci scalar in this theory. The general formulation for reconstructing the modified $f(R)$ gravity for any given FRW metric was developed by Nojiri and Odintsov \cite{Nojiri2006}. Capozziello et al. \cite{Capozziello2008} examined the exact solution of cosmological models in the context of $f(R)$ gravity. Nojiri and Odintsov \cite{Nojiri2008} proposed and demonstrated that the $f(R)$ gravity model as a viable candidate for unifying early-time inflation with late-time cosmic acceleration. In the review of $f(R)$ theories, Felice and Tsujikawa \cite{Felice2010} address various topics, including inflation, dark energy, cosmological perturbations, local gravity constraints, and spherically symmetric solutions in weak and strong gravitational backgrounds, etc. In $f(R)$ gravity, Tripathy and Mishra \cite{Tripathy2016} have described the dynamics of anisotropic locally rotationally symmetric (LRS) Bianchi type-$I$ models and obtained the solutions to the field equations. Recently, Agrawal et al. \cite{Agrawal2021} explored the Bianchi type-I cosmological model in the $f(R)$ theory of gravity with perfect fluid and discussed the behavior of gravitational baryogenesis using two distinct forms of the Ricci scalar.

Another method for explaining dark energy within the context of modified gravity is holographic dark energy (HDE) \cite{Li2004, Setare2007}, which has received a lot of interest in recent years. The EoS parameter is one of the most important cosmic parameters to describe the behavior and dynamics of the universe. The holographic principle is one of the fundamental quantum gravitational principles that may be used to investigate the nature of dark energy. Nojiri and Odintsov \cite{Nojiri2017} studied the generalized holographic dark energy spacetime in $ f(R) $ gravity with perfect fluid, where the corresponding infrared cut-off is determined by combining several FRW parameters. In general relativity, Sarkar and Mahanta\cite{Sarkar2013} and Santhi et al. \cite{Santhi2018} investigated the correspondence between HDE and quintessence cosmological models in various contexts.

Friedmann-Robertson-Walker (FRW) models are the best for representing the present large-scale structure of the universe. The FRW model is homogeneous and isotropic in nature. When we think about the early universe, we believe that it was not like today, i.e. it was not isotropic. To know the early structure of the universe, we should consider different models. In this sense, the Bianchi-type models are the simplest, most homogeneous, and most anisotropic; therefore, they are important to studying the beginning of the universe. In the Bianchi model, the spatial section is flat, where the extension or contraction rate is direction-dependent. Mete et al. \cite{Mete2013} obtained an analytic solution to the general relativistic field equation in four and five dimensions using a two-fluid Kasner-type Bianchi type-I cosmological model. The field equations of $f(R,T)$ gravity for Bianchi type-I space-time have been resolved by Adhav \cite{Adhav2012}. In different Bianchi type-I for the Kasner form metric, Saaidi et al. \cite{Saaidi2012} investigated the evolution of the Bianchi type-I Kasner metric in the context of $ f(R) $ gravity and observed that the power law $ f(R) $ models behaved like the quintessence model. Pawar and Dagwal \cite{Pawar2015} investigated the tilted Bianchi type-I Cosmological model of Kasner form in the context of the Brans-Dicke theory of gravity. Paliathanasis et al.\cite{Paliathanasis2018} showed the existence and stability of the Bianchi type-I Kasner solution in the context of $ f(T) $ gravity theory field equations using power and exponential laws.

The cosmic strings are one-dimensional topological defects that are necessarily formed during phase transitions in the early universe and, therefore, must be understood in terms of their evolution and observable traces. The galaxies are believed to have originated in the early universe due to density fluctuations caused by strings. Adhav et al. \cite{Adhav2009} explored the Bianchi type-III cosmological model with cosmic strings and domain walls in the context of general relativity theory. In addition, Harko and Matthew \cite{Harko2015} investigated Kasner-type static, cylindrically symmetric spacetime to obtain string solutions in the $f(R,L m)$ theory of gravity. Aditya and Reddy \cite{Aditya2018} showed that the universe achieves isotropy at late times by investigating the LRS Bianchi type-I spacetime with cosmic strings in the $ f(R) $ gravity. In $ f(R) $ gravity, Santhi et al. \cite{Santhi2019} have recently studied the LRS Bianchi type-I bulk viscous string cosmological model, which shows that the realistic energy conditions are satisfied. Moreover, Chirde et al. \cite{Chirde2020} explored the Bianchi type-I perfect fluid string cosmological model using $ f(T) $ gravity and found that quintessence dark energy dominates the universe.

As a result of the previous discussion, we investigate Bianchi type-I Kasner space-time in the context of the $f(R)$ theory of gravity in the present research. In the presence of string and holographic dark energy, the field equations of $f(R)$ gravity for the Kasner metric have been found. The geometrical and physical aspects of the models are also being explored. A discussion and conclusion are included in the concluding part. In section \ref{basic_eq}, basic equations are presented. Sections \ref{model_I} and \ref{model_II} are devoted to solutions of $ f(R) $ are gravity field equations. Section \ref{conclusion} summarizes the obtained results.

\section{Basic equations}\label{basic_eq}
Kasner-type spacetime is the most fundamental example of homogeneous and anisotropic spacetimes. They have been extensively used to analyze and classify cosmological singularities in the context of general relativity \cite{Belinskii1970}. Deruelle \cite{Deruelle1989} considered these metrics in the Gauss-Bonnet case a long time ago. We consider the Kasner type Bianchi type-I space-time is given by
\begin{equation}\label{Eq:1}
	ds^2=dt^2-t^{2P_1}dx^2-t^{2P_2}dy^2-t^{2P_3}dz^2
\end{equation}
where $P_1,P_2,P_3$ are three parameters satisfying the relation as $P_1+P_2+P_3=S$ and   $P_1^2+P_2^2+P_3^2=Q$. Brevik and Pettersen \cite{Brevik1997} investigated the consequences of the equation of state for cosmic fluids, including shear and bulk viscosities, using the Bianchi type-I metric of the Kasner form. Moreover, Cataldo and Campo \cite{Cataldo2000} showed that a viscous Bianchi type-I cosmological model of the Kasner form cannot describe the evolution of entropy in the universe.

The simplest modification of the general theory of relativity is the $ f(R) $ theory of gravity, and its Einstein-Hilbert action is modified by replacing the Ricci scalar $ R $ with the arbitrary function $ f(R) $ in the following form:
\begin{equation}\label{Eq:2}
	S=\frac{1}{2k^2}\int d^4x\sqrt{-g}f(R)+S_m\left(g_{\mu\nu},\psi \right) 
\end{equation}
where $k^2=8\pi G,\psi$ refers collectively to all matter fields,  $ f(R) $ is a general function of Ricci scalar $ R $. The field equations are obtained by varying the action \eqref{Eq:2} with respect to the metric $ g_{\mu\nu} $ as
\begin{equation}\label{Eq:3}
	F(R)R_{\mu\nu}-\frac{1}{2}f(R)g_{\mu\nu}-\bigtriangledown_{\mu}\bigtriangledown_{\nu}F(R)+g_{\mu\nu}\Box F(R)=k^2T_{\mu\nu}
\end{equation}
where $\Box F(R)=\frac{1}{\sqrt{-g}}\partial_{\mu}\left(\sqrt{-g}g^{\mu\nu}\partial_{\mu} \right), \Box=\bigtriangledown^{\mu}\bigtriangledown_{\mu} $. The energy momentum tensor of string and HDE is taken as
\begin{equation}\label{Eq:4}
	T_{\mu\nu}= T^s_{\mu\nu}+T^h_{\mu\nu}
\end{equation}
where $ T^s_{\mu\nu}=\rho_{s}u_{\mu}u_{\nu}-\lambda x_{\mu}x_{\nu}$ and $ T^h_{\mu\nu}=\left(\rho_{h}+p_{h}\right)u_{\mu}u_{\nu}- g_{\mu\nu}p_{h} $ with $ u_{\mu}u^{\mu} =-x_{\mu}x^{\mu}=1$ and $ u_{\mu}x^{\mu} =0$.
$ \rho_{s}=\rho_{p}+\lambda $ - denotes the rest energy density for a cloud of strings with particle attached to them, $ \rho_{p} $ - the density of particles, $\lambda$ - the cloud string tension density, $x^{\mu}$ - the direction of strings, $u^{\mu}$ - the four velocity vector, $\rho_{h}$ and $p_{h}$ the energy density and pressure of HDE respectively. It should be not here that the attempt of combination of string and HDE will provide some food for thought to discuss the interacting effect of topological defect and HDE. The primary object of the paper is to arouse the interest in the subject and to arrive at serviceable truth. The second approach of the subject is regarding theoretical background of string and HDE has moved us to put before you some of the conclusion which may be support the subject.

The standard energy conditions corresponding to cosmic string can be stated as \cite{Letelier1979}
\begin{itemize}
	\item[i)] Weak energy conditions: \qquad $ \rho_{s}\ge \lambda $ with $ \lambda\ge0 $,
	\item[ii)] Strong energy conditions: \qquad $ \rho_{s}\ge 0 $ with $ \lambda<0 $,
	\item[iii)] Dominant energy conditions: \quad $ \rho_{s}\ge 0 $ with $ \rho_{s}^2\ge \lambda^2 .$
\end{itemize}
Now, the field equations \eqref{Eq:3} for the metric \eqref{Eq:1} and energy momentum tensor \eqref{Eq:4} reduces to
\begin{equation}\label{Eq:5} 
	F(R)\left(1-S\right)P_{1} t^{-2}+\frac{1}{2}f(R)+\left(P_{1}-S\right)t^{-1}F_{4}-F_{44} =k^{2}P_{h} , 
\end{equation} 
\begin{equation}\label{Eq:6} 
	F(R)\left(1-S\right)P_{2} t^{-2}+\frac{1}{2}f(R)+\left(P_{2}-S\right)t^{-1}F_{4}-F_{44} =k^{2}\left(P_{h}-\lambda\right), 
\end{equation}
\begin{equation}\label{Eq:7} 
	F(R)\left(1-S\right)P_{3} t^{-2}+\frac{1}{2}f(R)+\left(P_{3}-S\right)t^{-1}F_{4}-F_{44} =k^{2}P_{h},
\end{equation}
\begin{equation}\label{Eq:8} 
	F(R)\left(Q-S\right)t^{-2}-\frac{1}{2}f(R)+SF_{4}t^{-1} =k^{2}\left(\rho_{s}+\rho_{h}\right) 
\end{equation}
where the suffix '4' denotes differentiation with respect to time.
From the equations \eqref{Eq:5} and \eqref{Eq:7}, we get
\begin{equation}\label{Eq:9}
	P_{1}=P_{3}
\end{equation}
We investigate two different forms of HDE, which are found in the literature. The first form of HDE considered by Granda and Oliveros \cite{Granda2008} as
\begin{equation}\label{Eq:10}
	\rho_{h}=3M_{p}^{2}\left(\alpha H^{2}+\beta \dot{H}\right)
\end{equation}
where $ \alpha, \beta $ are constants, $ H=\frac{\dot{a}}{a} $ is Hubble parameter, $ a $ is scale factor and $ M_{p} $ is the reduced plank mass with $ M_{p}^{-2}=1 $. The second form of  HDE is considered by Saadat \cite{Saadat2011} as
\begin{equation}\label{Eq:11}
	\rho_{h}=3d^{2}H^{2}
\end{equation}
where $ d $ is constant.\
The fields equations \eqref{Eq:5}-\eqref{Eq:8} are four non-linear equations in five unknowns $ F $, $ p_{h} $ $\rho_{h}$, $\rho_{s}$ and $\lambda$. To obtain the exact solutions, we require one extra condition, which we consider in the following sections (3) and (4):

\section{Model-I}\label{model_I}
In this section, we assume a relation between $F$ and scale factor $a$. Johri and Desikan \cite{Johri1994} have considered the power law relation between the average scale-factor $a(t)$ and scalar function $ \phi $ in Brans-Dicke theory. Recently, in the context of $f(R)$ theory, Uddin et al.\cite{Uddin2007} has examined the result by taking $F \propto a^{i} $, where $i$ - is an arbitrary integer. Following to Johri and Desikan \cite{Johri1994} and Uddin et al.\cite{Uddin2007}, we assume the power law relation between $F$ and $a$ as 
\begin{equation}\label{Eq:12}
	F=la^{m}=lt^{mS}
\end{equation}
where $ a=\sqrt{-g}=t^{S} $. Equation \eqref{Eq:12} further gives us
\begin{equation}\label{Eq:13}
	f(R)=l\left(S^{2}-2S+Q\right)t^{mS-2}
\end{equation}
where $R=\left(S^{2}-2S+Q\right)t^{-2}$ and $ l $ is integrating constant.
Solving equations \eqref{Eq:5}, \eqref{Eq:6} and with the help of equations \eqref{Eq:12} and \eqref{Eq:13}, we obtain the string tension density as
\begin{equation}\label{Eq:14}
	\lambda=\frac{\left(P_{1}-P_{2}\right)(1-S+mS) l t^{mS-2}}{k^{2}}
\end{equation}
\begin{figure}[ht]
	\centering
	\includegraphics[width=0.7\linewidth]{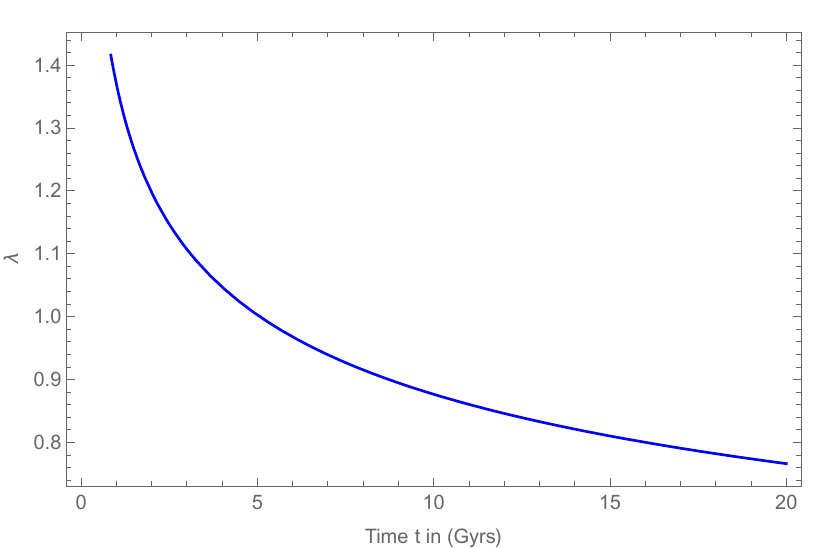}
	\caption{Plot of string density $\lambda$ \textit{vs} time $ t $}
	\label{fig:1}
\end{figure}

Now, The density of HDE is obtained by using equations \eqref{Eq:10},\eqref{Eq:11} and \eqref{Eq:13} as
\begin{equation}\label{Eq:15}
	\rho_{h}=\frac{3d^{2}S^{2}}{t^{2}}
\end{equation}
Also, solving equations \eqref{Eq:5}, \eqref{Eq:13} and \eqref{Eq:14}, the pressure of HDE ($ p_{h} $) and rest energy density ($\rho_{s}$) are obtain as
\begin{equation}\label{Eq:16}
	p_{h}=\frac{P_{1}(1-S)(S^2-2S+Q)+(P_{1}-S)mS-mS(mS-1)l t^{mS-2}}{k^{2}}
\end{equation}
\begin{equation}\label{Eq:17}
	\rho_{s}=\frac{-3k^{2}d^{2}S^{2}t^{-2}+\left(\frac{Q-S^{2}}{2}+mS^{2}\right)lt^{mS-2}}{k^{2}}
\end{equation}
Again, The density of HDE and string in the form of Granda and Oliveros \cite{Granda2008} is obtained by using equations \eqref{Eq:10},\eqref{Eq:12} and \eqref{Eq:13} as
\begin{equation}\label{Eq:18}
	\rho_{h}=3M_{p}^{2}\left(\frac{\alpha S^{2}-\beta S}{t^{2}}\right)
\end{equation}
\begin{equation}\label{Eq:19}
	\rho_{s}=\frac{-3k^{2}M_{p}^{2}(\alpha S^{2}-\beta S)t^{-2}+\left(\frac{Q-S^{2}}{2}+mS^{2}\right)lt^{mS-2}}{k^{2}}
\end{equation}
\begin{figure}[ht]
	\centering
	\includegraphics[width=0.7\linewidth]{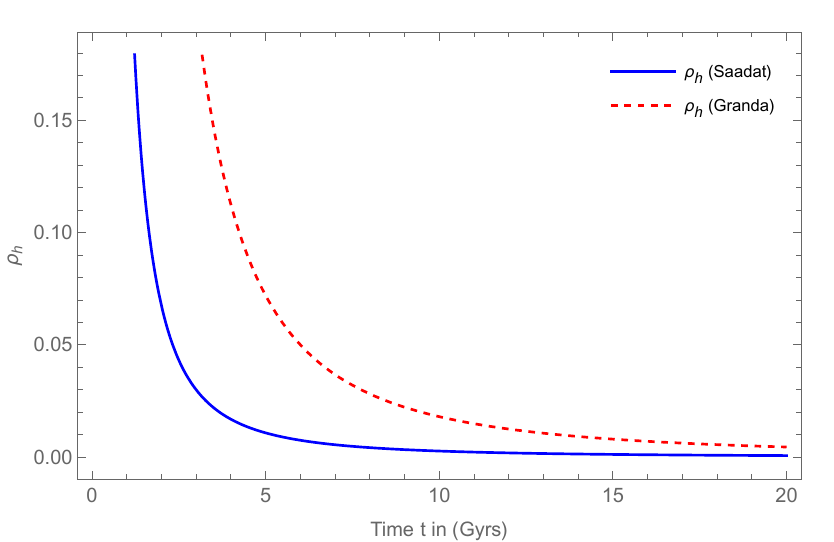}
	\caption{Plot of Holographic dark energy $\rho_h$ \textit{vs} time $ t $}
	\label{fig:2}
\end{figure}

From Figure \ref{fig:2}, it is observed that the HDE density is a decreasing function of time $t$. It is large near $t=0$ and tends to zero as $t \rightarrow \infty$. The behavior of HDE density is the same for Granda and Oliveros \cite{Granda2008}, and Saadat \cite{Saadat2011} form. It should be noted here that the variation of $\rho_h$ is strictly different than that of Bianchi type-I metric in $f(R)$ are gravity \cite{Katore2020}.
\begin{figure}[ht]
	\centering
	\includegraphics[width=0.7\linewidth]{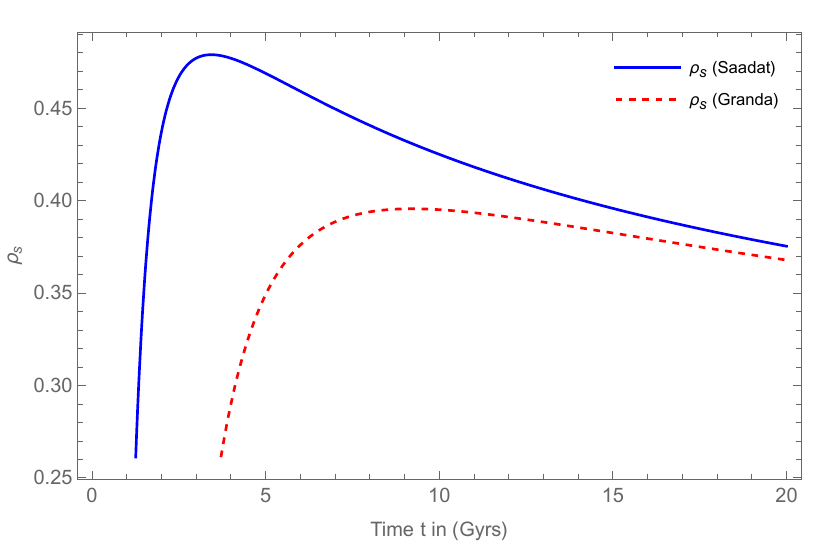}
	\caption{Plot of rest energy density for a cloud strings with particle $\rho_s$ \textit{vs} time $ t $}
	\label{fig:3}
\end{figure}

From figure \ref{fig:3}, it is clear that the rest energy density of string is positive. It is zero near $ t=0 $ which suddenly increase to its maximum value and decreases with increasing time and constant as $ t \rightarrow \infty$. The effect of HDE for two different types i.e. Saadat \cite{Saadat2011} and  Granda and Oliveros \cite{Granda2008} form is shown in figure \ref{fig:3}.
Also, from figure \ref{fig:1} and \ref{fig:3} it is clear that $ \lambda \ge0 $, $ \rho_s \ge0 $ and $ \lambda \ge \rho_s $ i.e. violation of weak energy condition. The positive value of string tension density means presence of string phase of the universe and negative value of $ \lambda $ indicates that string phase of the universe is disappear i.e. universal dominated by cosmological constant \cite{Martinez2005}. Here $ \lambda \ge0 $, therefore, the model is in favour of presence of string phase of the universe.
\begin{figure}[ht]
	\centering
	\includegraphics[width=0.7\linewidth]{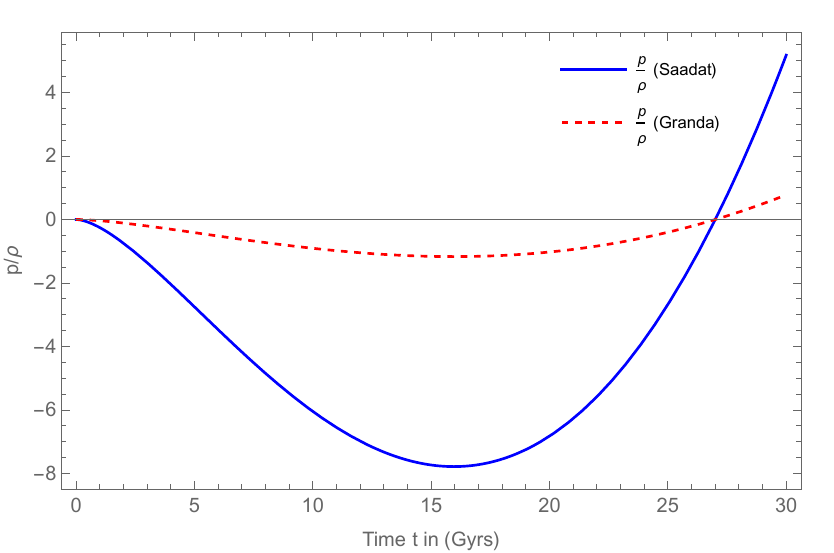}
	\caption{Plot of $p/\rho$ \textit{vs} time $ t $}
	\label{fig:4}
\end{figure}

The plot of $p/\rho$ is shown in figure \ref{Eq:4}. We found that $p/\rho =0$ near $ t=0 $ i.e. the universe was dominated by dust matter, and as $ t $ increases to the present ($ t\ge 14 $ Gyrs), $p/\rho$ is negative and $0 \le p/\rho \le -1$ for Saadat \cite{Saadat2011} type HDE whereas $0 \le p/\rho \le -8$ for Granda and Oliveros \cite{Granda2008} types HDE. In the future ($ t=14 $ Gyrs), $p/\rho$ tends to be positive and become greater than 1, i.e., the universe may be dominated by ekpyrotic type \cite{Lehners2008} of HDE. According to SNIa observational data, $-1.67 < p/\rho < -0.62$ \cite{Knop2003} whereas combination of SNIa data, CMB anisotropy, and galaxy clusters studies placed the limit $-1.33 < p/\rho < -0.79$ \cite{Tegmark2004b}. In this case, it is important to note that the range of $p/\rho$ for Saadat \cite{Saadat2011} type of HDE is more appropriate to observational data.

The important physical parameters for the given metric: The expansion scalar $(\theta)$ of the model is obtained as
\begin{equation}\label{20} 
	\theta =3H=\frac{3(2P_{1}+P_{2})}{t} , 
\end{equation}
From equation \eqref{20}, we see that the expansion scalar $\theta$ is a decreasing function of time. Therefore the expansion rate was large near $ t=0 $, and as $t \rightarrow \infty$, it tends to zero.\
The deceleration parameter $ q $ is determine as
\begin{equation}\label{21} 
	q=\frac{d}{dt} \left(\frac{1}{H} \right)-1=\frac{1}{2P_{1}+P_{2}}-1
\end{equation}
The sign of deceleration parameter $ q $ indicates whether the universe is accelerating or decelerating. A positive sign of $ q $ indicates a decelerating universe and a negative sign of $ q $ indicates an accelerating universe. From expression \eqref{21}, it is clear that $ q $ is positive for $ 1 > 2P_{1}+P_{2} $ and negative for $ 1 < 2P_{1}+P_{2} $ i.e. the universe is accelerating for $ 1 < 2P_{1}+P_{2} $, which clearly depends on the parameters $ P_{1} $, $P_{2} $ of Kasner metric \eqref{Eq:1}.

\section{Model-II}\label{model_II}
In this case, we assume the functional form proposed by Starobinsky \cite{Starobinsky2007} in the form $ f(R)=R+bR^{n} $, which is widely discussed in $ f(R) $ theory of gravitation and leads to the accelerated expansion of the universe. Here, we consider the model given as
\begin{equation}\label{Eq:22}
	f(R)=R+bR^{\eta}
\end{equation}
where $b>0$ and $\eta>0$. Using equations \eqref{Eq:22}, \eqref{Eq:5} and \eqref{Eq:6}, we get
\begin{equation}\label{Eq:23}
	\lambda=\frac{(P_{1}-P_{2})(1-S)}{k^{2}t^{2}}+\frac{\eta b(P_{1}-P_{2})(3-S-2\eta)(S^{2}-2S+Q)^{\eta-1}}{k^{2}t^{2\eta}}
\end{equation}
\begin{figure}[ht]
	\centering
	\includegraphics[width=0.7\linewidth]{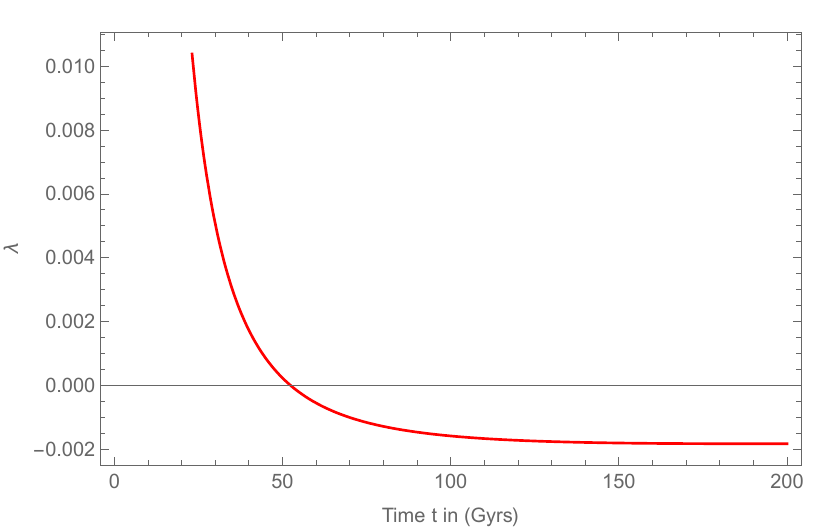}
	\caption{Plot of string density $\lambda$ \textit{vs} time $ t $}
	\label{fig:5}
\end{figure}
From figure \ref{fig:5}, it is clear that the string tension density $\lambda$ is positive decreasing function of time $t$. Near $ t=0 $, it is large and as $t \rightarrow \infty$, $\lambda \rightarrow -0.002$ i.e. the model is in favor string phase of universe at early stage of the universe which later on disappear.
Using equations \eqref{Eq:22} and \eqref{Eq:5}, we obtain
\begin{equation}\label{Eq:24}
	\begin{split}
		p_{h}=&\frac{\left(P_{1}\eta(1-S)+\frac{S^{2}-2S+Q}{2}+(P_{1}-S+2\eta-1)\eta(2-2\eta)\right)b (S^{2}-2S+Q)^{\eta-1}}{k^{2}t^{2\eta}}\\
		&+\frac{P_{1}(1-S)+\frac{S^{2}-2S+Q}{2}}{k^{2}t^{2}}
	\end{split}
\end{equation}
Using equations \eqref{Eq:8}, \eqref{Eq:10}, \eqref{Eq:11} and \eqref{Eq:22}, the rest energy density of string are obtained as
\begin{equation}\label{Eq:25}
	\rho_{s}=\frac{(\frac{Q-S^2}{2})-3k^{2}d^{2}S^{2}}{k^{2}t^{2}}+\frac{\left(\eta(Q-S)-\frac{(S^{2}-2S+Q)}{2}+\eta S(2-2\eta)\right) b(S^{2}-2S+Q)^{\eta-1}}{k^{2}t^{2\eta}}
\end{equation}
\begin{equation}\label{Eq:26}
	\begin{split}
		\rho_{s}=&\frac{\left(\eta(Q-S)-\frac{(S^{2}-2S+Q)}{2}+\eta S(2-2\eta)\right) b(S^{2}-2S+Q)^{\eta-1}}{k^{2}t^{2\eta}}\\
	&+\frac{(\frac{Q-S^2}{2})-3k^{2}M_{p}^{2}(\alpha S^{2}-\beta S)}{k^{2}t^{2}}
	\end{split}
\end{equation}
\begin{figure}[ht]
	\centering
	\includegraphics[width=0.7\linewidth]{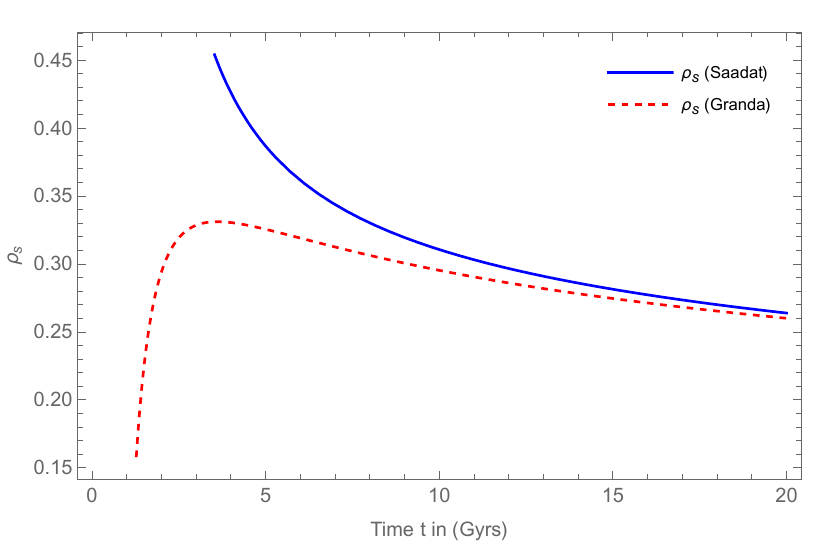}
	\caption{Plot of rest energy density $\rho_s$ \textit{vs} time $ t $}
	\label{fig:6}
\end{figure}
\begin{figure}[ht]
	\centering
	\includegraphics[width=0.7\linewidth]{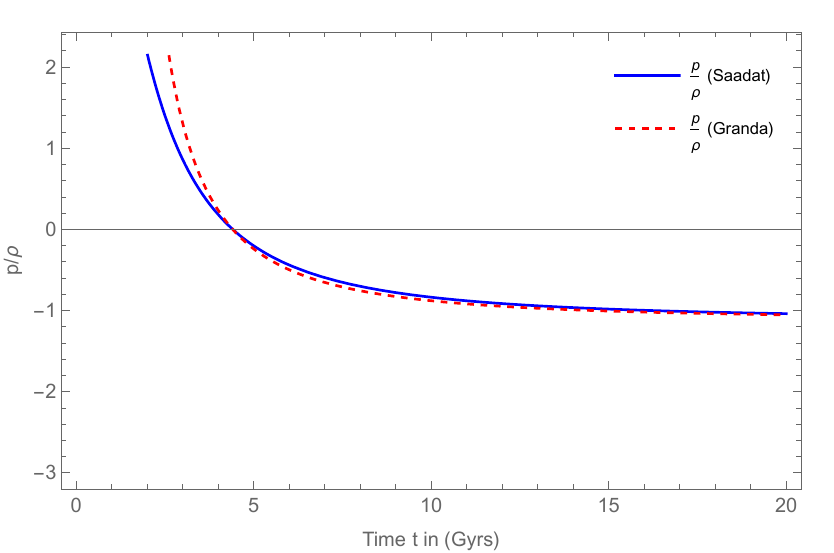}
	\caption{Plot of $p/\rho$ \textit{vs} time $ t $}
	\label{fig:7}
\end{figure}

From figure \ref{fig:7}, it is clear that $p/\rho  >0$ near $ t=0 $ and $p/\rho \rightarrow -1 $  as $ t \rightarrow \infty$, i.e., the universe was dominated by the ekpyrotic type of HDE and at present, it is dominated by $ \Lambda $CDM type HDE. The behavior of $p/\rho$ for Saadat \cite{Saadat2011} type HDE and Granda and Oliveros \cite{Granda2008} type HDE is similar.
The novelty of the work is that effect of $ f(R) $ function on different terms like $ \lambda $, $ \rho_s $ and $p/\rho$ is clearly observed from figures \ref{fig:1} and \ref{fig:5}, \ref{fig:3} and \ref{fig:6}, \ref{fig:4} and \ref{fig:7} respectively. Further, in both the models $ \lambda \ge 0$, which was not seen in our earlier studies in modified $ f(G) $ \cite{Katore2018} and $ f(T) $ \cite{Chirde2020} gravity. In the case of $ f(R,T) $ theory of gravitation, the existence of string phase may depend on the term $ R^m $ of $ f(R,T) $ function \cite{Hatkar2018}.

\section{Conclusion}\label{conclusion}
In the present paper, we have explored Kasner space-time in the presence of holographic dark energy coupled cloud strings in the context of the $ f(R) $ theory of gravitation. We have found the following results:
\begin{itemize}
	\item In model-I: Figure \ref{fig:2} shows that the energy density remains positive throughout the evolution of the universe. String phase of the universe is present at early stage of evolution of the universe.	The universe is dominated by quintessence type holographic dark energy at present, which may later turn into ekpyrotic type of HDE.
	\item In model-II: The $ \rho_s $ behaves differently for Saadat \cite{Saadat2011} type HDE and Granda and Oliveros \cite{Granda2008} types HDE. The study confirms the presence of string phase of the universe in the early stage of evolution of the universe. Moreover, we seen that $p/\rho \rightarrow -1 $ as $t \rightarrow \infty$ i.e. the universe may dominated by $ \Lambda $CDM type HDE at present. The rest energy density of string is positive which satisfy the dominant energy conditions $\rho_{s} \ge 0$ and $\rho_{s}^2 \ge \lambda^2$. The behaviour of HDE density for Saadat \cite{Saadat2011} and Granda and Oliveros \cite{Granda2008} type form is different. In case of Saadat \cite{Saadat2011} type, it is decreasing function of time, initially (near $ t=0 $) it is large and as $ t \rightarrow \infty$, $ \rho_s $ tends to constant. In case of Granda and Oliveros \cite{Granda2008} type, it starts with constant value at $ t=0 $ and attains its maximum and tends to constant value with increasing time.
	\item The universe is accelerating for $ 1 < 2P_{1}+P_{2} $. It is expending and rate of expansion is decreasing function of time.
	\item In this consideration of string and HDE even though they are non-interacting, the effect of HDE on the term $\rho_s $ is observed. It could be very interesting to consider the interacting model of string and HDE.
\end{itemize}

\end{document}